# Time-resolved study of carbonization and growth of ultrathin 3C-SiC on Si(111) under ultra-high vacuum

Pranjali Jadhao[1], Mojdeh Fallahpour[1], Josef Polčák[1,2], Eva Kolíbalová[1], Michal Horák[1], Jan Michalička[1] Petr Bábor[1,2], Stanislav Voborný[1,2] and Tomáš Šikola[1,2,*]

[1]*Central European Institute of Technology, Brno University of Technology, Purkyňova 656/123, Brno 612 00, Czech Republic*

[2]*Institute of Physical Engineering, Brno University of Technology, Technická 2896/2, Brno 616 69, Czech Republic*

**e-mail:* sikola@fme.vutbr.cz

## Abstract

The early-stage formation of silicon carbide (SiC) on Si(111) by ethylene exposure under ultra-high vacuum (UHV) was investigated to resolve the time-dependent chemical and morphological evolution of an ultrathin layer. Clean Si(111) substrates were exposed to $C_2H_4$ at 800 °C for 2 min to 4 h, and the surfaces followed by in-situ X-ray photoelectron spectroscopy (XPS) and ex-situ AFM, SEM, AES, SIMS and TEM. Si 2p and C 1s peak analysis shows the progressive conversion of elemental silicon into a carbidic Si–C phase, the SiC fraction overtaking the elemental component between 120 and 160 min and saturating near 80–81% beyond 180 min, leaving about 19–20% residual elemental silicon. Correlative SEM and AFM reveal a parallel morphological progression, from sparse isolated islands to a coalesced, near-continuous layer. AES depth profiling confirms carbon incorporated into the near-surface region rather than weakly adsorbed as contamination, assigning the islands to early SiC nuclei. TEM confirms the zinc-blende lattice and the presence of cubic silicon carbide (3C-SiC). Together, these results provide a time-resolved picture of SiC nucleation, coalescence and layer growth on Si(111), relevant to 3C-SiC heteroepitaxy, and can be utilized in the optimization of SiC/Si(111) templates for growth of III-nitride and other carbide systems on silicon (e.g. $Mo_2C$).

## 1. Introduction

Silicon carbide (SiC) has emerged as one of the foremost wide-bandgap semiconductors for power and high-frequency electronics, combining a high breakdown field, a high electron drift velocity, and excellent thermal conductivity [1,2]. Of its many polytypes, the cubic form (3C-SiC) is the only one with a zinc-blende structure, and it is particularly attractive for transistors and integrated electronics because of its high carrier mobility and comparatively low density of interface states [3]. The polytypes differ chiefly in the hexagonality of their stacking sequence, which tunes the bandgap over a remarkably wide range, from about 2.3 eV for the cubic phase to 3.3 eV for the 2H phase [4]. A further advantage of 3C-SiC is its stability at relatively low deposition temperatures (below ~1500 °C), which makes it amenable to growth on foreign substrates [5]. The central obstacle to exploiting 3C-SiC is the absence of an affordable native substrate, which forces its growth by heteroepitaxy, most commonly on silicon [3,6]. Silicon is, however, a poorly matched host: its lattice parameter differs from that of 3C-SiC by roughly 20%, and its thermal expansion coefficient by about 8% at room temperature [4]. These mismatches generate large residual stresses and a characteristic family of extended defects (misfit dislocations, stacking faults, microtwins, and antiphase boundaries) that degrade the epilayer [7]. To contain them, heteroepitaxy conventionally begins with a carbonization step: the heated silicon surface is exposed to a carbon-bearing precursor alone, with no separate silicon source, so that the outermost silicon atoms are converted in place into a thin SiC buffer that seeds the subsequent growth [5]. Since this buffer templates everything that follows, its structural quality largely dictates the defect density and morphology of the final film, making carbonization one of the most decisive stages of the entire process [8].

Until now, the most of the studies connected with the preparation of a SiC buffer layer have rested almost entirely on Si(100). The equivalent process on Si(111) is far less well understood, despite the strong technological pull of that orientation: its surface atomic arrangement mirrors the (0001) basal plane of the wurtzite nitrides, so SiC on Si(111) serves as a standard buffer for the heteroepitaxy of GaN and related III-nitrides [9]. The few dedicated Si(111) studies have relied on different precursors or growth environments, for example acetylene at low-pressure chemical vapour deposition [10] or carbon monoxide in solid–gas-phase epitaxy [11], and have shown that substrate orientation strongly influences void shape, grain coherency, and even the crystallographic orientation of the resulting SiC [6,11]. Yet these reports concentrate largely on reaction conditions, growth windows, and final-film quality; the sequence by which SiC islands nucleate, coalesce, and relax into a continuous epilayer during the earliest instants on Si(111) has received comparatively little attention [12]. The time evolution of early stages of the SiC buffer layer formed on Si(111) under controlled ethylene exposure therefore remains largely undocumented.

One can also come across the need for the SiC buffer layer during the growth of other carbidic materials than SiC. In attempting to grow two-dimensional molybdenum carbide ($Mo_2C$) [13] on Si(111) by depositing molybdenum in an ethylene gas atmosphere under ultra-high vacuum conditions, we consistently found that molybdenum silicide rather than the intended carbide was formed for substrate temperatures of 900–1250 °C. This finding has directed our attention to the formation of a SiC buffer layer acting also possibly as a diffusion barrier of C towards the Si(111) substrate, and thus to the early stages of carbonization examined here.

Accordingly, the present work tracks the formation of SiC on Si(111) by ethylene exposure in ultra-high vacuum for growth times spanning from 2 min to 4 h, a range chosen so that the slow reaction kinetics can be resolved stage by stage. The evolving surfaces were monitored in situ by X-ray photoelectron spectroscopy (XPS) and characterised ex situ by scanning electron microscopy (SEM), atomic force microscopy (AFM), Auger electron spectroscopy (AES), time-of-flight secondary ion mass spectrometry (TOF-SIMS) and scanning transmission electron microscopy (STEM) with electron energy-loss spectroscopy (EELS). Placed on a common time axis, these measurements allow us to trace the Volmer–Weber Island growth in time, observing the chemical-state evolution and the nucleation-to-coalescence sequence, the interface structure and buffer-layer development together with the appearance of interfacial voids. Beyond its immediate bearing on 3C-SiC heteroepitaxy and on SiC/Si (111) templates for the III-nitrides [9], this study forms a useful platform for growth of $Mo_2C$, and related carbides, on the Si (111) surface.

## 2. Experimental details

The experiments were performed in a complex ultra-high-vacuum (UHV) apparatus [14] that allows Si and SiC structures to be analysed in situ by XPS without exposing them to the ambient atmosphere. XPS was carried out using a setup comprising a DAR400 X-ray source and an EA125 hemispherical electrostatic analyser (both Omicron). Measurements were performed at room temperature with Al Kα radiation, and the pressure during measurement was always better than $4.5 \times 10^{-7}$ Pa. The surface morphology was examined ex situ by SEM (FEI Verios 460L) and AFM (Bruker Dimension ICON). Site-specific AES was performed with a Scienta Omicron nanoSAM instrument, combined with Ar+ sputtering at 2 keV ion energy, 10 mA emission current, and 2 µA sample current, with the beam rastered over the analysed region in two successive 20 min cycles. Depth-resolved elemental analysis was carried out by TOF-SIMS (ION-TOF, TOF SIMS5). A TEM lamella was prepared using a gallium-focused ion beam (FIB) in a dual-beam FIB/SEM system (Helios NanoLab 660, FEI). High-resolution and analytical scanning transmission electron microscopy (STEM) investigations were performed using a Cs-image- and Cs-probe-corrected TITAN Themis 60-300 microscope (Thermo Fisher Scientific, USA) operated at 300 kV. Electron energy-loss spectroscopy (EELS) was carried out using a GIF Continuum spectrometer equipped with a STELA camera (Gatan, USA). EELS elemental maps were acquired in Single EELS mode with an energy dispersion of 3.0 eV $channel^{-1}$ using the following parameters: beam current of 50 pA, dwell time of 0.2 ms $pixel^{-1}$ and step size of 0.36 nm. EELS data acquisition and analysis were performed using Gatan Microscopy Suite (GMS, Gatan). The elemental maps presented were generated from individual ionization edges (O-K, Mo-L, Si-K, and C-K) after background subtraction.

Samples of $5 \times 18$ $mm^2$ were cut from an n-type, phosphorus-doped Si (111) wafer (resistivity 0.01–0.02 Ω·cm). They were cleaned ultrasonically in acetone and then in isopropyl alcohol (IPA) for 5 min each, rinsed in deionised water, and dried under a nitrogen stream. The samples were subsequently transferred into the UHV apparatus (base pressure $3 \times 10^{-8}$ Pa) and annealed by direct resistive heating at 800 °C overnight to remove the native oxide and surface contamination prior to growth. The sample temperature was measured with an infrared radiation pyrometer, and the removal of surface contaminants was confirmed in situ by XPS.

SiC growth was carried out in an UHV chamber (base pressure in the low $10^{-8}$ Pa range). Ethylene ($C_2H_4$, 99.95%; Messer) was introduced through a precision leak valve while the substrate was held at 800 °C, a temperature selected to ensure complete dissociation of the chemisorbed ethylene and the desorption of hydrogen, and to promote the formation of crystalline SiC. The resulting working pressure was 2.5 × $10^{-4}$ Pa as measured by an ion gauge. Growth durations of 2, 5, 10, 30, 50, 80, 120, 150, and 240 min were used to follow the time evolution of the film.

## 3. Results and discussion

### 3.1. X-ray photoelectron spectroscopy

The surface chemistry during ethylene exposure was followed by in-situ XPS, monitoring the Si 2p and C 1s core-level peaks. Only silicon and carbon were detected at every stage; no oxygen or other species were present, so the surface evolves within a clean Si–C system. After annealing without exposure to gas (*t* = 0 min), Si 2p spectral region [Fig. 1(a)] shows a single elemental Si peak at 99.12 eV. After 40 min exposure to $C_2H_4$, a carbidic Si–C component appears at 100.42 eV. The elemental component weakens with time. The Si–C component dominates the Si 2p envelope by 240 min. The 1.30 eV shift between the two states is consistent with the Si–C chemical shifts reported for thin SiC layers on Si [6,15] (see Table S1 in the Supplementary Information (SI)).

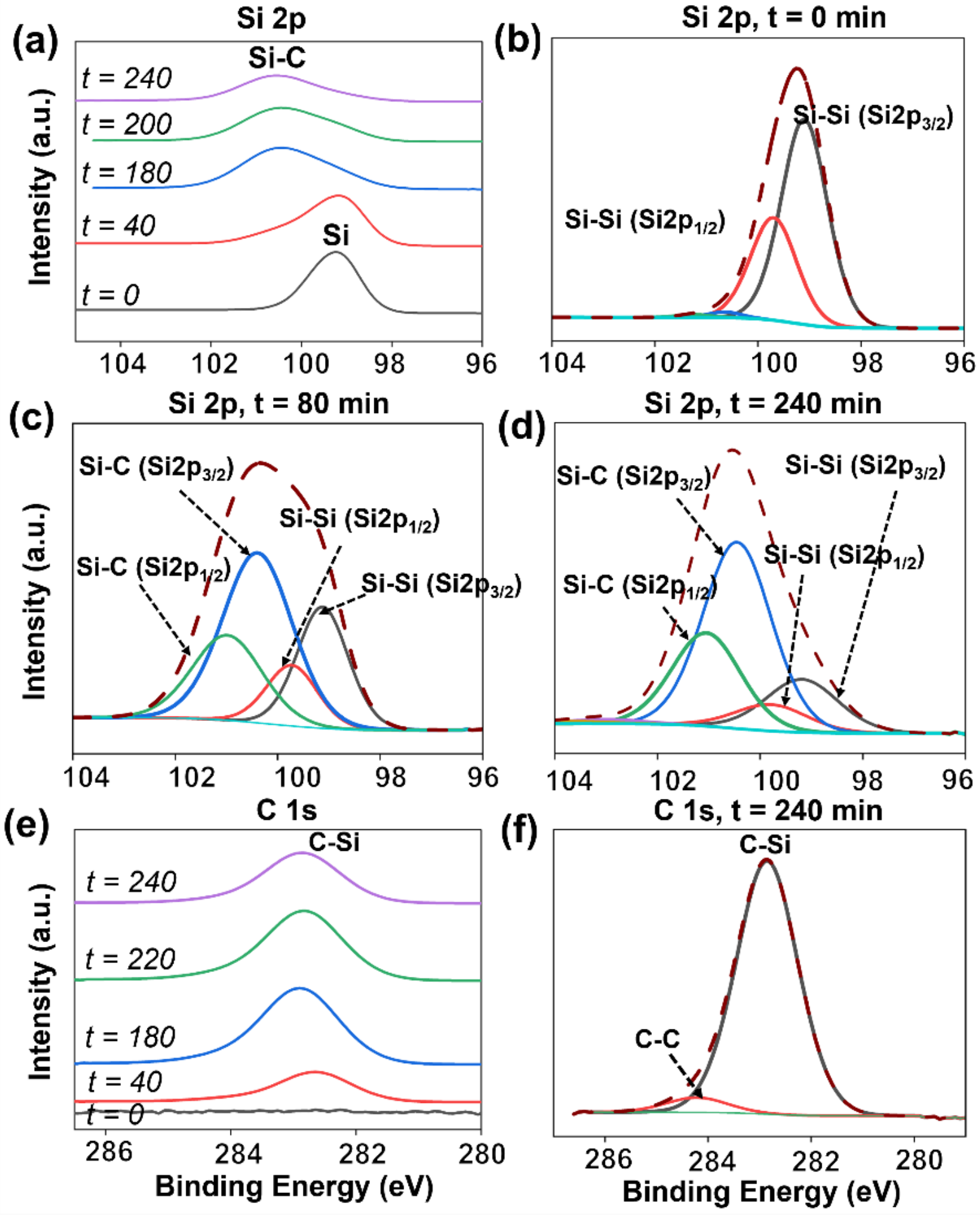

***Fig. 1.*** *Si 2p and C 1s XPS spectra during SiC growth on Si(111) at 800 °C. (a) Si 2p waterfall spectra for selected $C_2H_4$ exposure times (in min), (b–d) Fitted Si 2p spectra at t = 0, 80, and 240 min. (e) C 1s waterfall at selected exposure time (in min), (f) Fitted C 1s spectrum for t = 240 min (carbidic C–Si at 282.85 eV, minor C–C at 284.3 eV).*

The fitted Si 2p spectra [Fig. 1(b–d)] resolve this conversion. The annealed surface [Fig. 1(b)] is characterised by a single elemental doublet with no carbide intensity. After $t$ = 80 min [Fig. 1(c)] a distinct Si–C doublet has appeared alongside the elemental one, and at $t$ = 240 min [Fig. 1(d)] the Si–C doublet is dominant, with the elemental Si reduced to a shoulder. Each spectrum was fitted with GL(30) line shapes and a fixed 2:1 doublet ratio (0.61 eV spin–orbit splitting, 0.61 eV FWHM) [16], so the integrated Si–C area provides a direct measure of the carbide fraction.

The carbon core level confirms the same trend. The C 1s waterfall [Fig. 1(e)] shows the carbidic peak absent for the annealed surface at $t$ = 0 min, and growing steadily with exposure at $t$ = 240 min, tracking the Si–C development in the Si 2p region. The fitted C 1s spectrum at $t$ = 240 min [Fig. 1(f)] is dominated by a single carbidic C–Si component at 282.85 eV, with only a weak C–C feature at a higher binding energy of 284.3 eV.

Table 1 lists the fitted binding energies and assignments. Two silicon states are resolved: elemental silicon, identified by the Si–Si component at 99.12 eV, and carbidic silicon, identified jointly by the Si–C (100.42 eV) and C–Si (283.0 eV) components. A minor C–C component (284.3 eV) indicates remnants of hydrocarbon species. All values lie within the reported literature ranges for silicon carbide, supporting the assignment of the carbide features to a stoichiometric SiC phase. Since the Si 2p and C 1s core-level binding energies identify the carbidic chemical state but are essentially insensitive to SiC polytypes, the cubic polytype (3C–SiC) is inferred from the growth conditions: SiC formed heteroepitaxially on Si substrates at these temperatures, including by carbonization, is known to adopt the 3C form [5,17]. This is directly confirmed by STEM/EELS analysis in Section 3.4.

**Table 1. Core-level binding energies and bond (chemical-state) assignments.**

| Element | Core level | Chemical state | BE (eV) | Assignment | Ref. |
|---|---|---|---|---|---|
| Si | Si $2p_{3/2}$ | Elemental $Si^0$ | 99.12 | Si–Si (Si) | [17] |
| | Si $2p_{3/2}$ | Silicon carbide | 100.42 | Si–C (SiC) | [17,18] |
| C | C 1s | Silicon carbide | 282.85 (~283) | C–Si (SiC) | [17,19] |
| | C 1s | C–C / hydrocarbon | 284.3 | $sp^2/sp^3$ C (minor) | [18,19] |

*Literature BE ranges: $Si^0$ 99.1–99.5 eV [17]; Si–C (SiC) 100.4–101.3 eV [6,15,18,20]; C–Si (SiC) 282.6–283.5 eV [18,19,20].*

The deconvoluted peak-component areas were used to quantify the surface composition as a function of exposure time. Figure 2(a) plots the relative fractions of the elemental Si and SiC phases in the modified layer, calculated from the areas of the Si–Si and Si–C components of the fitted Si 2p doublet (both spin–orbit components summed). Similarly, Fig. 2(b) shows the exposure-time dependence of the relative fractions of the SiC and hydrocarbon-derived phases, calculated from the areas of the C–Si and C–C components of the fitted C 1s peak. The corresponding peak components, with their binding energies, relative areas, and chemical-state assignments for each exposure, are summarised in Table 2; the Si–C relative areas tabulated there are identical to the SiC fractions plotted in Fig. 2.

**Table 2. XPS peak components, binding energies, relative areas, and chemical states.**

| Exposure time (in min) | Peak component | BE (eV) | Relative area* (%) | Chemical state |
|---|---|---|---|---|
| **0** | Si–Si | 99.12 | 100.0 | Elemental Si |
| | Si–C | – | n.d. | Silicon carbide |
| | C–Si | – | n.d. | Silicon carbide |
| | C–C | – | n.d. | Adventitious carbon |
| **40** | Si–Si | 99.12 | 85.2 | Elemental Si |
| | Si–C | 100.42 | 14.8 | Silicon carbide |
| | C–Si | 282.64 | 90.7 | Silicon carbide |
| | C–C | 284.00 | 9.3 | Adventitious carbon |
| **80** | Si–Si | 99.12 | 73.7 | Elemental Si |
| | Si–C | 100.41 | 26.3 | Silicon carbide |
| | C–Si | 282.96 | 92.1 | Silicon carbide |
| | C–C | 284.32 | 7.9 | Adventitious carbon |
| **220** | Si–Si | 99.12 | 18.9 | Elemental Si |
| | Si–C | 100.42 | 81.1 | Silicon carbide |
| | C–Si | 282.82 | 93.3 | Silicon carbide |
| | C–C | 284.30 | 6.7 | Adventitious carbon |
| **240** | Si–Si | 99.12 | 19.2 | Elemental Si |
| | Si–C | 100.42 | 80.8 | Silicon carbide |
| | C–Si | 282.85 | 94.5 | Silicon carbide |
| | C–C | 284.21 | 5.5 | Adventitious carbon |

**Values from CasaXPS fits (Si 2p doublet: 0.61 eV splitting, 2:1 ratio, 0.6 eV FWHM, GL(30)). Si–Si + Si–C and C–Si + C–C each sum to 100% within their respective regions; n.d. = not detected.*

The fitted spectra reveal a continuous conversion of elemental silicon into its carbide. The Si–C fraction increases from below 15% at $t$ = 40 min, through 26% at $t$ = 80 min, to 81% at $t$ = 220 min, where it saturates (80.8% at 240 min), while the elemental Si fraction falls correspondingly from 85% to 19%. Figure 2(a) shows these two phases crossing near 120–160 min, the point at which SiC becomes the majority surface phase, and levelling beyond 180 min at a SiC fraction of about 80–81%. The residual elemental Si of about 19% at saturation indicates either that the conversion is not completed throughout the whole layer detectable by XPS (for example, owing to negligible diffusion of C atoms from, and of Si atoms towards, the surface through the already formed carbide layer) or this layer is not sufficiently continuous making the Si substrate beneath the layer for XPS visible. As follows from the sections below, the latter is the case.

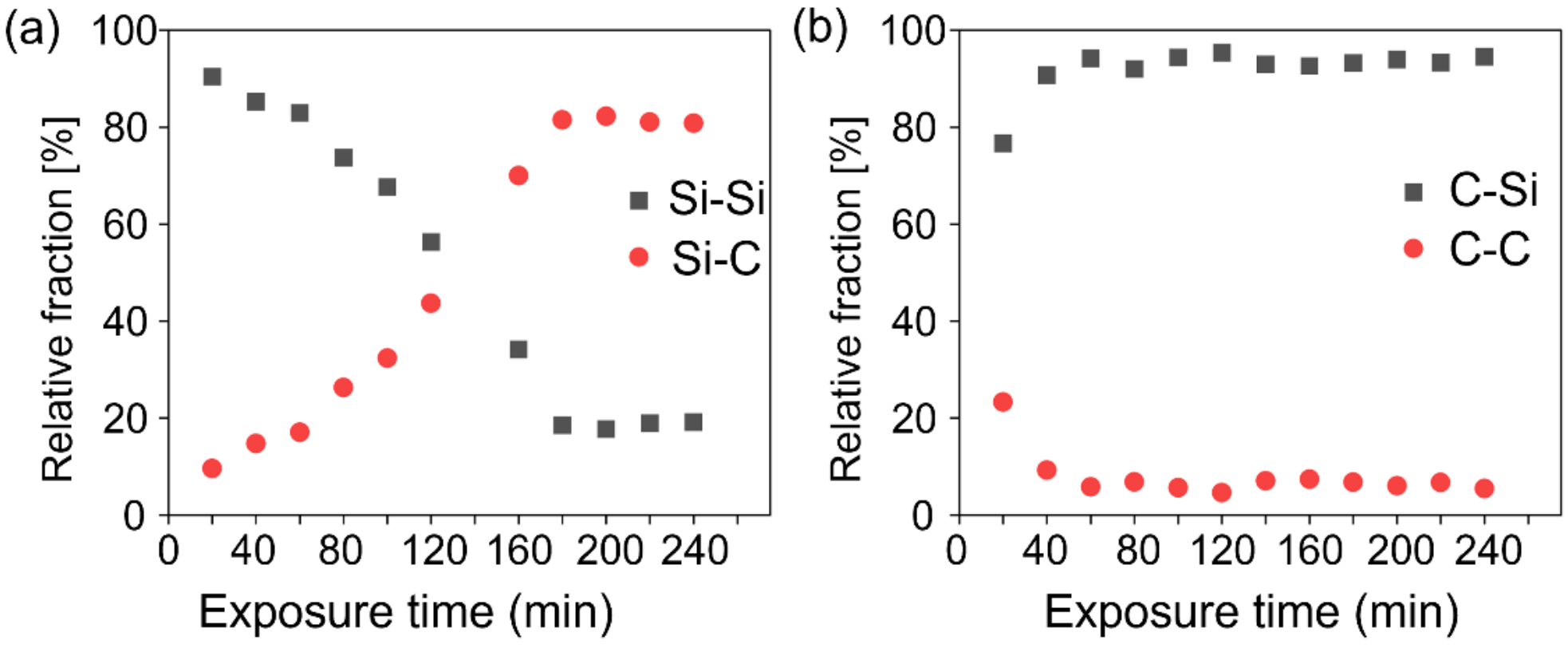

***Fig. 2**. (a) Relative fractions of the elemental Si and SiC phases in the modified layer as a function of exposure time, calculated from the areas of the Si–Si and Si–C components of the fitted Si 2p doublet (both spin–orbit components summed). (b) Relative fractions of the SiC and hydrocarbon-derived phases, calculated from the areas of the C–Si and C–C components of the fitted C 1s peak.*

On the carbon side, the C 1s signal is dominated by the C–Si component at all exposures, rising from 90.7% at $t$ = 40 min to 94.5% at $t$ = 240 min [Fig. 2(b)]. The C–C component is correspondingly minor and decreases steadily with exposure time, from about 9.3% at $t$ = 40 min to 5.5% at $t$ = 240 min, indicating that prolonged exposure progressively lowers the hydrocarbon-derived carbon content. This low and decreasing C–C fraction confirms minimal hydrocarbon contamination and a high-purity carbidic surface, with the carbon incorporated almost entirely as SiC. The agreement between the high C–Si fraction (>90%) from the C 1s analysis and the saturating Si–C fraction (~81%) from the Si 2p analysis provides internally consistent evidence for a well-defined SiC formation.

## 3.2. Surface morphology: correlative SEM and AFM

To follow the evolution of the surface morphology during SiC growth, correlative SEM and AFM measurements were carried out on the samples grown for $t$ = 2, 5, 10, 30, 50, 80, 150, and 240 min at 800 °C, complementing the chemical-state evolution described in Section 3.1.

The SEM [Fig. 3(a–h)] and AFM [Fig. 4(a–h)] images reveal a consistent morphological evolution that proceeds in two distinct stages. In the first stage (2–30 min), increasing exposure produces increasingly numerous and more closely spaced islands: the SiC nuclei grow in size, the distance between them shrinks, and the surface develops a denser arrangement of laterally growing islands, already evident as a near-complete island layer by 30 min. Annealing produces atomic terraces separated by step edges, which can act as preferential nucleation sites; step edges are known to template two-dimensional nucleation during growth on Si(111) [21], and SiC island formation on annealed, stepped Si(111) has been reported previously [22].In [Fig. 3 (a)] these terraces are obviously seen by the decoration of the SiC islands.

In the second stage (50–240 min), the trend reverses: the apparent lateral size of individual features decreases as coalescence proceeds, and the islands merge into a near-continuous layer. At the two longest exposures [Figs. 3, 4(g–h)], larger features with multi-layered structure and large deep voids begin to appear. The agreement between the SEM and AFM datasets confirm that this is a genuine morphological transition rather than a technique-specific artefact.

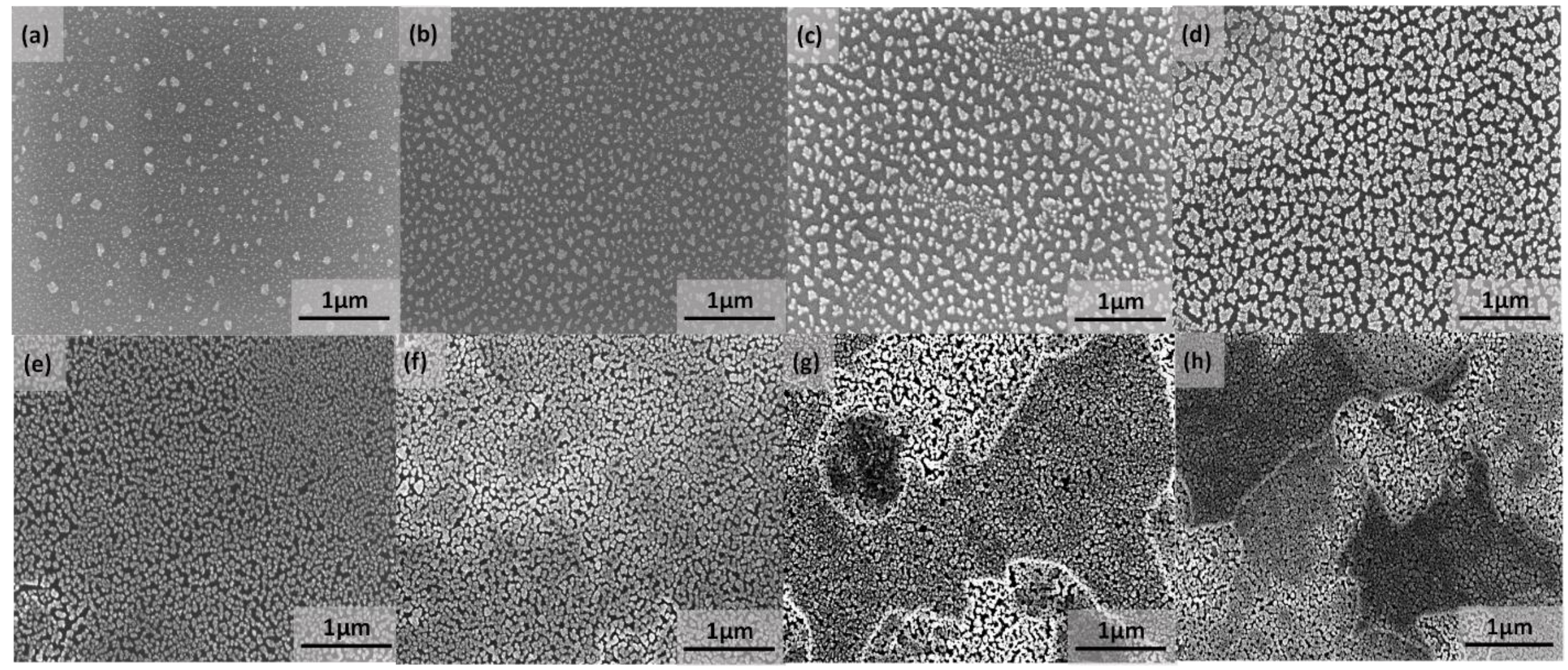

***Fig. 3***. *SEM images of of SiC nanostructures on Si (111) at (a) 2, (b) 5, (c) 10, (d) 30, (e) 50, (f) 80, (g) 150, and (h) 240 min of* $C_2H_4$ *exposure.*

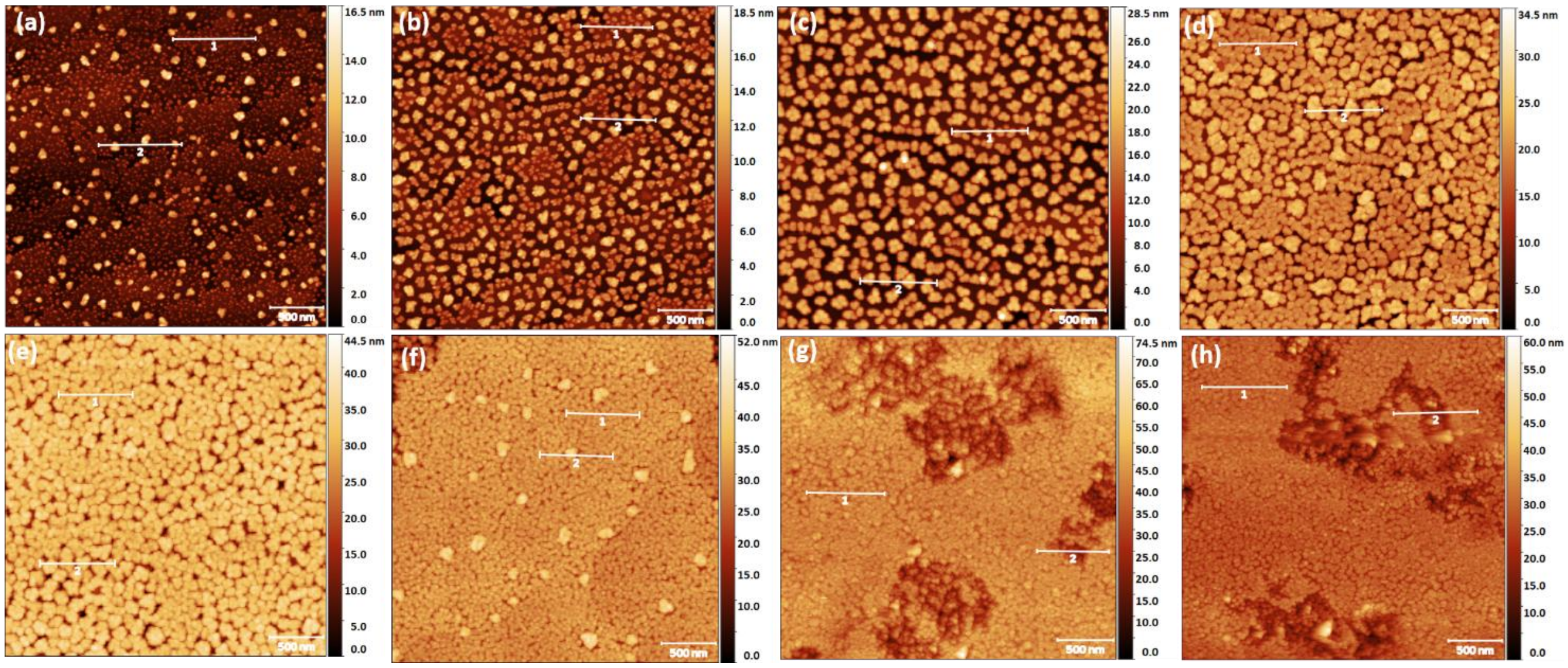

***Fig. 4***. *AFM images (3 × 3 µm scan area) of the same samples as in Fig. 3, after (a) 2, (b) 5, (c) 10, (d) 30, (e) 50, (f) 80, (g) 150, and (h) 240 min.*

The height profiles extracted from the AFM topography images [Fig. 4(a–h)] quantify this evolution [Fig. 5(a–h)]. Feature heights increase from about 1–8 nm at $t$ = 2 min to about 5–20 nm at $t$ = 30 min, consistent with continued vertical and lateral growth during nucleation. At later exposures, the height range narrows and drops, from about 3–17 nm at 50 min to about 3–4 nm at 240 min, reflecting the flattening that accompanies island coalescence. In addition to the flat areas in Figs 5(g) and 5(h) corresponding to exposure times of 150 and 240 min, respectively, there are also large deep voids inside the layer (~24 nm).

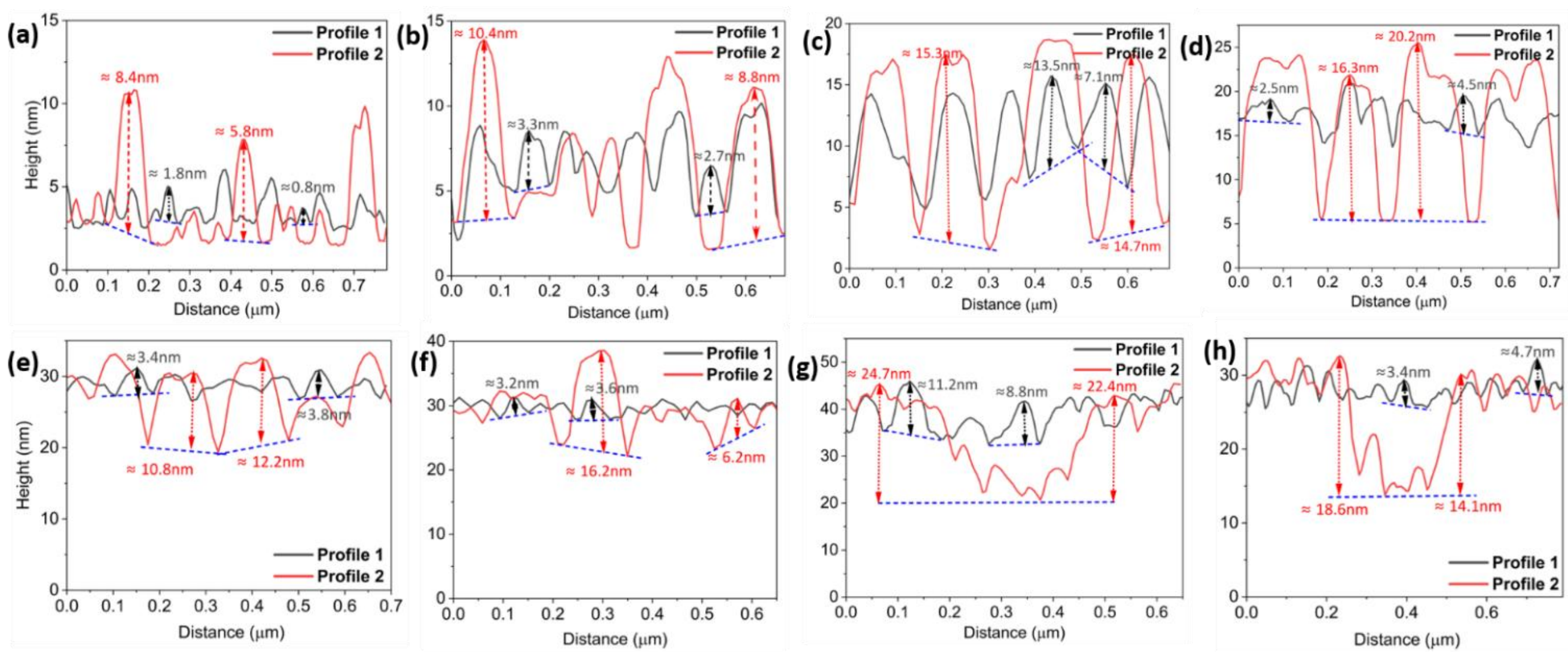

***Fig. 5.*** *Height profiles extracted along the lines marked in the AFM images in Fig. 4, after (a) 2, (b) 5, (c) 10, (d) 30, (e) 50, (f) 80, (g) 150, and (h) 240 min. Profile 2 (black curves) were taken from smoother Profile 1 (red curves) go through bigger islands.*

Hence, we conclude the silicon signal detected by XPS for these longest exposure times [Fig. 2] comes from the silicon surface at the bottom. Obviously, considering the possible growth of molybdenum carbides on this buffer layer, one should pay attention to the presence of these voids and try to minimize them. Probably, by a bit lower exposure time than those longest ones, for instance 80 – 100 min, when there is already a pronounced coalescence resulting in a quasi-continuous layer with smaller voids. In addition, the growth temperature should be kept as low as possible, to maintain the silicon diffusion towards the surface low. Hence, low-temperature growth methods delivering reactive carbon-based species like ultra-low energy ions (< 50 eV) or atoms/molecular fractions should be a preferential choice.

The RMS surface roughness [Fig. 6] rises sharply from about 2.1 nm at (2 min) to a maximum of 5.5 nm at (10 min), reflecting rapid, heterogeneous nucleation, then decreases steadily through 4.6 nm (30 min), 4.5 nm (50 min), 3.2 nm (80 min), and 3.3 nm (150 min), to about 2.5 nm at $t = 240$ min. This monotonic decline beyond the early maximum tracks the coalescence seen in Figs. 3 and 4, indicating a progressively more uniform surface despite the occasional larger voids visible at the two the longest exposures

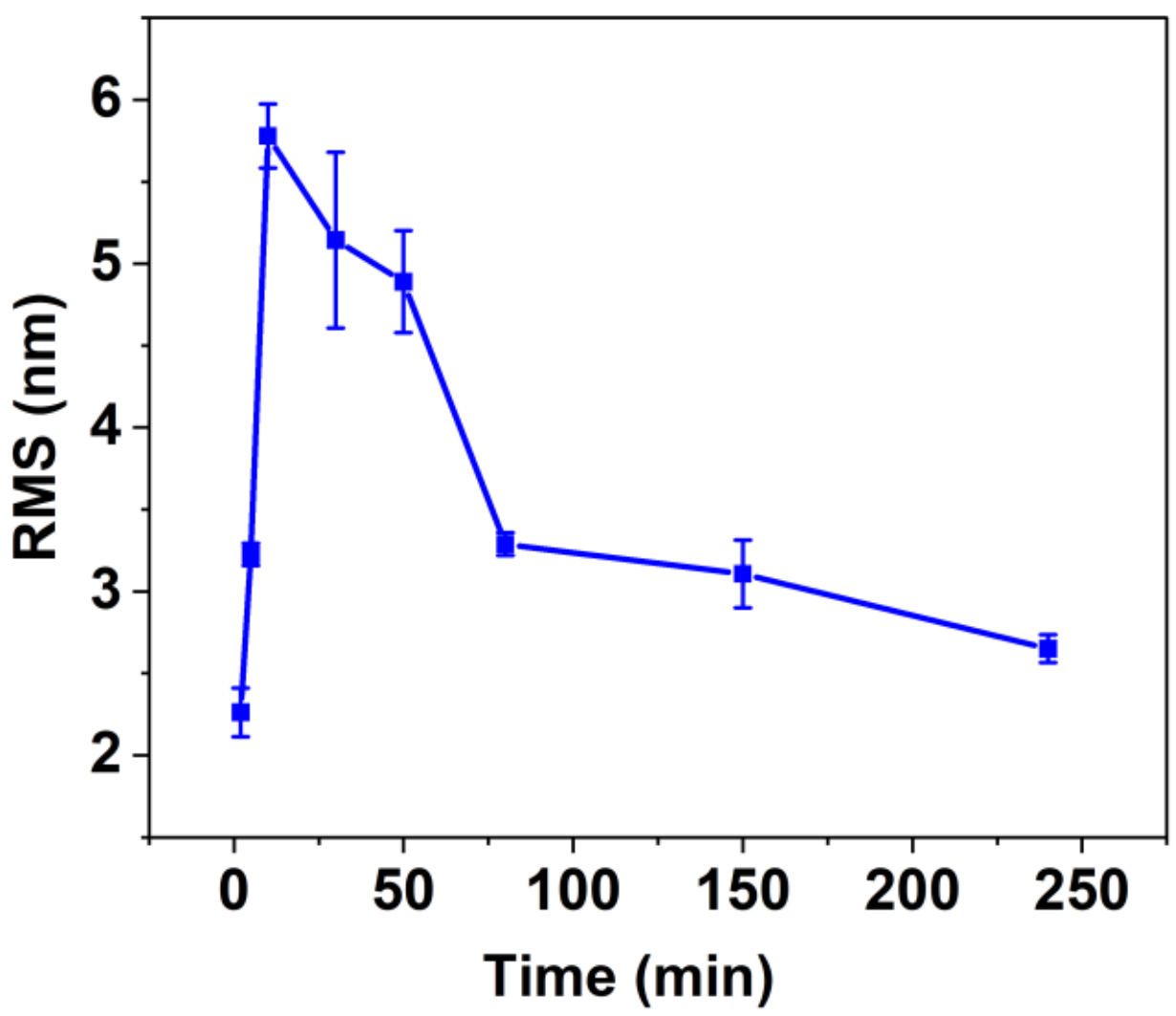

***Fig. 6.*** *RMS surface roughness versus exposure time, calculated from the AFM topography images in Fig. 4.*

### 3.3. Auger electron spectroscopy

Site-specific AES (lateral resolution better than 6 nm at 10 keV) was used to probe ex situ whether carbon at this growth stage is preferentially contained in the islands visible by SEM and AFM. Two locations on the 10 min sample were analysed: *S1*, a valley between islands, and *S2*, an island, before ion-beam sputtering and after each of two 20 min $Ar^+$ sputtering cycles [Fig. 7 (a)]. Survey spectra confirmed C, Si, and O as the only significant species at either location, with no other elemental contamination detected (Figs. S1 and S2 in the SI).

Differentiated C KLL and Si LMM spectra [Fig. 7(b and c)] show a comparable Si signal at both locations but markedly stronger carbon at S2, consistent with a localised rather than uniform carbon distribution. The C/Si peak-to-peak ratio was 0.97 at S2 versus 0.48 at S1 before sputtering, and this carbon enrichment remained significant throughout the sputtering sequence (0.66 vs 0.31 after 20 min; 0.56 vs 0.24 after 40 min; Fig. 7(d), Table S2); the S2/S1 ratio of these C/Si values in fact rose slightly, from 2.0 to 2.3, across the sequence. This contrasts with the oxygen signal which dropped sharply at both locations after the first sputter cycle - the effect consistent with the removal of an adventitious surface oxide. The persistence of the carbon enrichment at S2, unlike oxygen, indicates that carbon is not confined to a removable surface layer, supporting localised early-stage carbide nucleation at the island sites. This behaviour is consistent with a previous scanning-Auger-microscopy study in which SiC nanoparticles formed on a Si surface from contaminant carbon during UHV annealing were identified by their localised C KLL signal [23].

We emphasise that AES, as applied here, probes the elemental composition and chemical environment of the near-surface region through the C KLL and Si LMM transitions, but carries no long-range structural (diffraction) information. Although the C KLL line shape is known to discriminate carbidic from graphitic carbon [24], it is not sensitive to the crystallinity or polytype of the carbide. Accordingly, these results establish localised carbide-phase nucleation at the island sites, while the identification of the carbide as 3C-SiC rests on the growth conditions (Section 3.1) and would require diffraction-based confirmation, for example by TEM, to be considered definitive (see below).

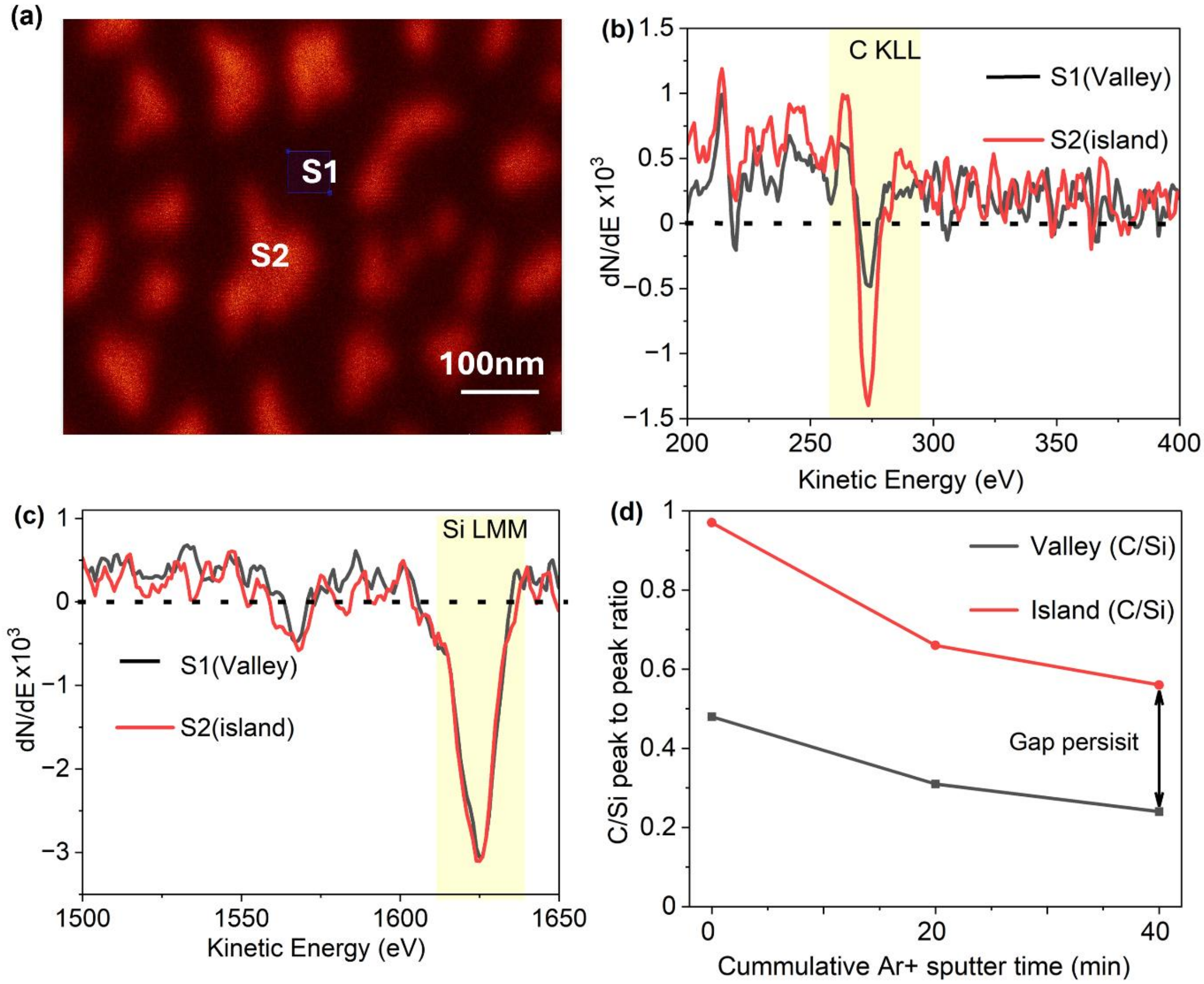

***Fig. 7***. *Site-specific AES analysis of the Si(111) sample exposed for 10 min to $C_2H_4$, at a spot in a valley between islands (S1) and at an early-stage island (S2). (a) SEM image of the region analysed by AES with the S1 and S2 spots marked (b) Differentiated (dN/dE) C KLL spectra of S1 and S2 after sputtering cycle (20 min $Ar^+$ sputtering), showing a substantially stronger carbon peak at S2. (c) Differentiated Si LMM spectra of S1 and S2, showing comparable silicon signals at both locations. (d) C/Si peak-to-peak ratio versus cumulative $Ar^+$ sputter time, showing the persistent carbon enrichment at S2 with respect to S1 throughout the sputtering sequence.*

### 3.4 SIMS depth profiling & TEM cross sectioning

To probe the depth distribution of carbon after prolonged growth, SIMS depth profiling was performed on the 240 min sample using an $O_2^+$ ion beam (2 keV, 620 nm) for sputtering and pulsed $Bi^+$ ion beam (30 keV, 5 pA) for analysis [Fig. 8]. The analysed area was 100 × 100 nm$^2$. The $C^+$ signal rises to a broad maximum within the first few seconds of sputtering, attributed to the SiC-containing surface layer, and subsequently decays by more than one order of magnitude without a sharp cutoff. The $^{30}Si^+$ signal is suppressed within the carbon-rich region and saturates at a steady level thereafter, showing that the profile terminates in the Si substrate. The initial seconds of both traces fall within the pre-equilibrium transient region of SIMS.

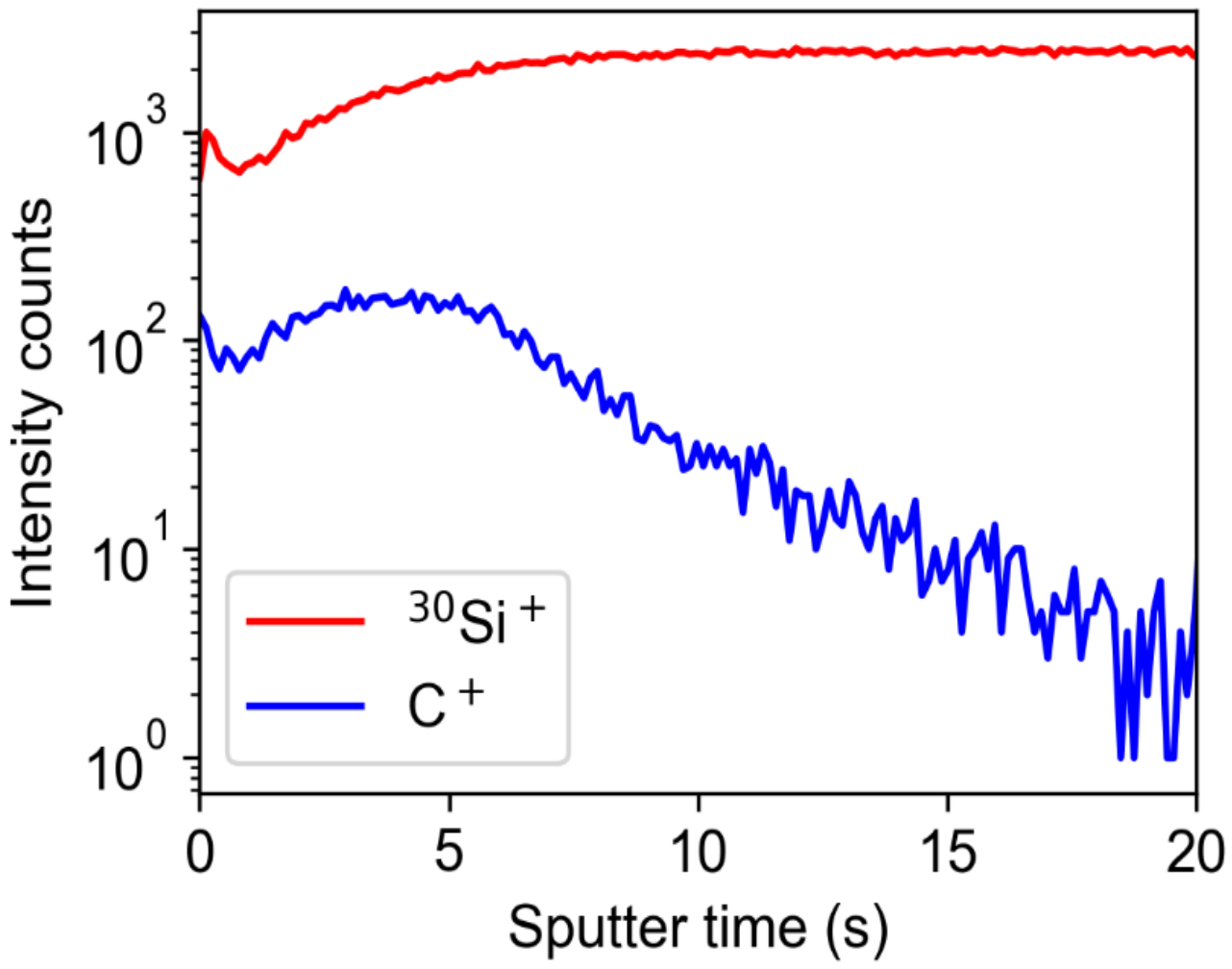

***Fig. 8.*** *SIMS depth profile of the 240 min sample showing the $^{30}Si^+$ (red) and $C^+$ (blue) secondary ion intensities versus sputter time on a logarithmic intensity scale. The $C^+$ maximum corresponds to the SiC-containing surface layer; its gradual decay reflects a broad SiC/Si transition consistent with the rough, island-derived interface and ion-beam mixing.*

This breadth is consistent with the island-derived morphology established by AFM and SEM (Section 3.2): a laterally and vertically non-uniform layer of coalesced three-dimensional islands, together with the interface roughening expected from Si out-diffusion during carbonisation [25–27], causes the sputter front to exit the SiC phase at different depths across the analysed area. Kinetic roughening of the growth surface is likewise a recognised feature of 3C-SiC heteroepitaxy on Si(111) [28]. This idea is supported by a cross section of the grown layer and its interface with the Si substrate obtained by STEM measurements on a lamella made from the sample by a Ga FIB (Fig. 9). In addition, ion-beam mixing and matrix effects associated with $O_2^+$ bombardment contribute to a profile broadening as well. Considering the erosion rate ~1.55 nm/s, the width of the $C^+$ maximum signal (~ 7s) in Fig. 8 corresponds to a depth of 11 nm. This is in reasonable agreement with the ~9 nm SiC layer thickness measured directly by STEM (Fig. 10).

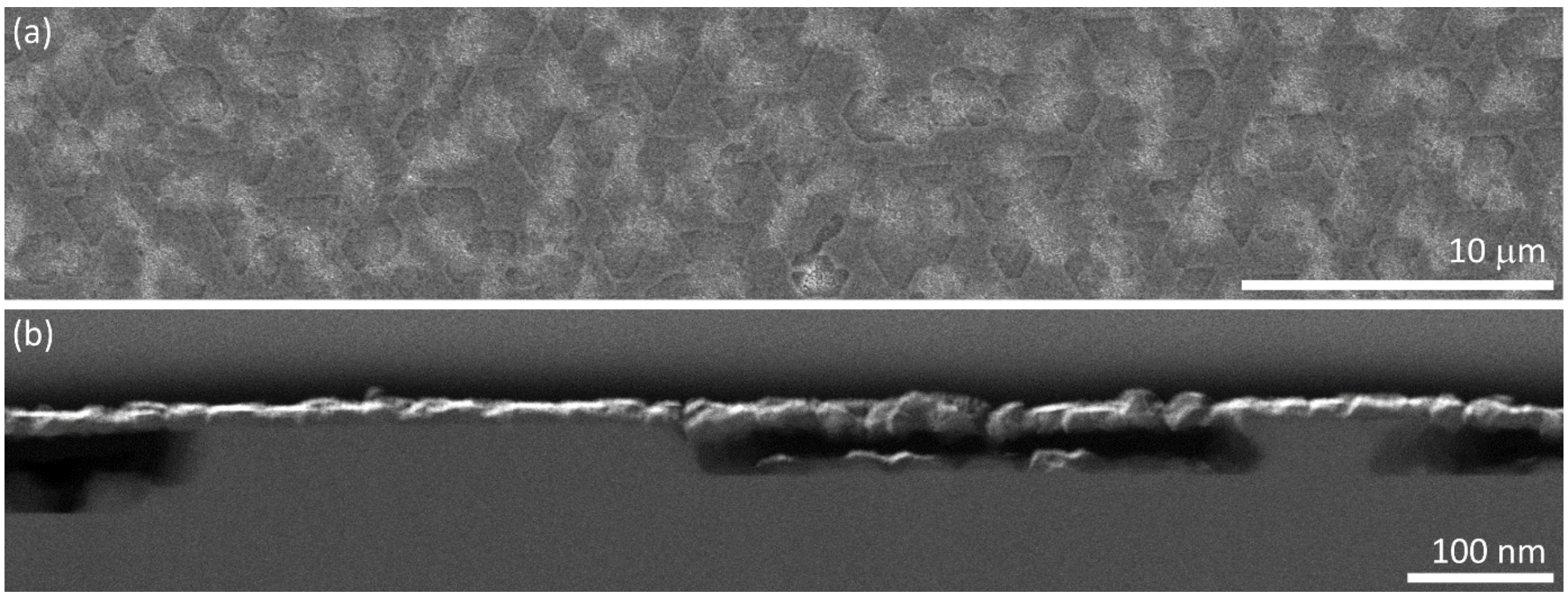

***Fig. 9.*** *Overview of the SiC layer grown for 240 min on the Si (111) substrate covered in situ by a Mo capping layer to prevent mixing of the carbon from the SiC and C protective layer and possible oxidation. (a) SEM image of the surface. (b) High-angle annular dark field (HAADF) STEM micrograph of a lamella prepared by Ga-FIB.*

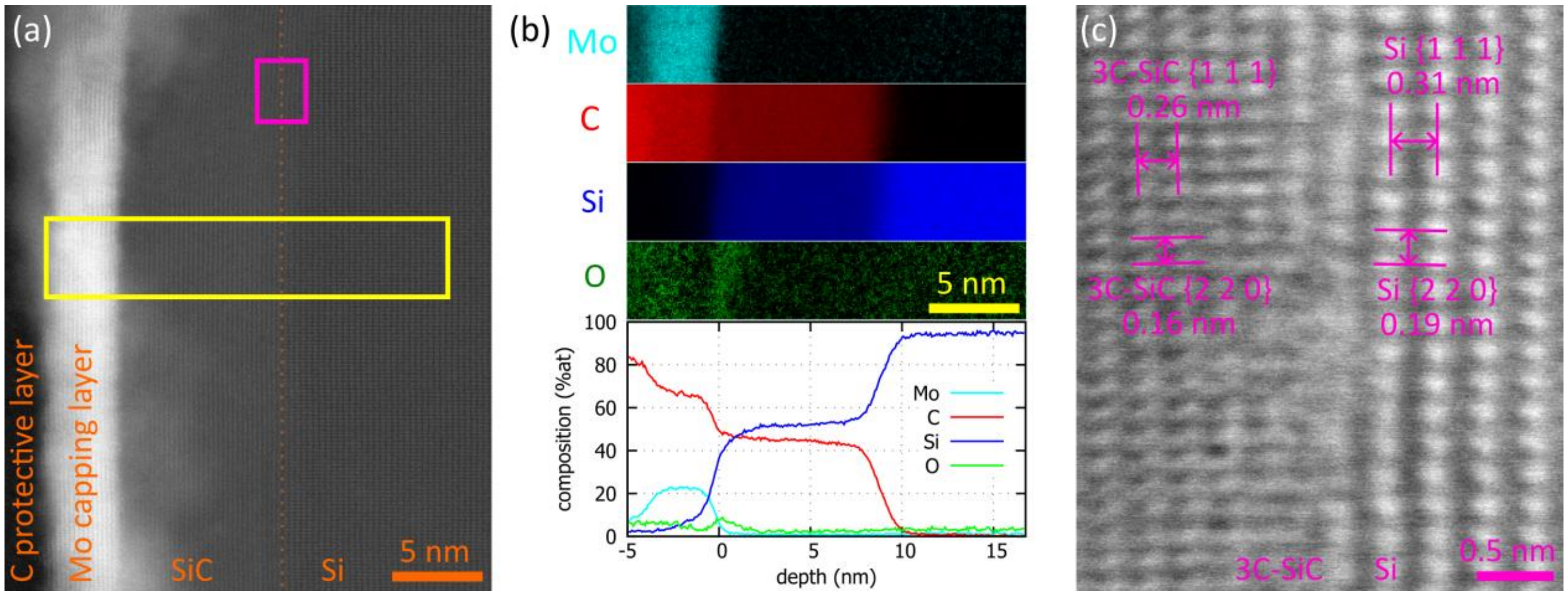

***Fig. 10.*** *Cross section of a homogeneous part of the 240 min SiC layer selected in an island – grain and investigated by STEM EELS. (a) Annular dark field (ADF) STEM image. (b) Elemental maps and profile of the SiC layer and its surroundings within the yellow rectangle marked in (a) provided by EELS showing the coexistence of Si and C elements roughly at 50 % at composition over the depth from 0 to 9 nm. (c) High resolution ADF STEM image of the interfacial SiC-Si area corresponding to a pink rectangle in (a).*

To get a more detailed information on the depth profile of the elemental composition and atomic structure of the 240 min SiC layer, STEM EELS experiments have been carried out on the lamella mentioned above. Importantly, before making the lamella, the SiC layer was covered in situ by a Mo capping layer (evaporated at RT) to prevent mixing of the carbon from the SiC and C protective layer and possible oxidation. The cross-section investigation of a homogeneous part selected in an island – grain of the layer is depicted in Fig. 10. The STEM image of the area [Fig. 10 (a)] shows a crystalline structure both in the SiC layer and Si substrate. Interestingly, it is also the case for the Mo capping layer. Consequently, sharp interfaces between the Mo and SiC layers, as well as between the SiC layer and the Si substrate are noticeable.

The depth profile is in detail analysed in Fig. 10(b) where the yellow rectangle marked in Fig. 10 (a) was inspected and the zero depth was assigned to the Mo – SiC interface. Hence, one can clearly determine the thickness of the SiC layer around 9 nm. This is also supported by the elemental distribution measured by EELS and showing the coexistence of Si and C elements roughly at 50 % of atomic composition over the depth from 0 to 9 nm. At a closure view, a transition area of these elements at the upper (depth ~ 0 nm) and lower (depth ~ 9 nm) interfaces are evident. An atomically resolved detail of the SiC – Si interface is shown in Fig. 10 (c) where the transition area spans over a few single layers, thus achieving thickness of ~ 1 nm.

Figure 10 (c) further reveals the interplanar distances perpendicular to the SiC – Si interface of 2.60 Å in SiC, corresponding to {1 1 1} planes of 3C-SiC, and 3.14 Å in Si, corresponding to {1 1 1} planes of Si. The interplanar distances parallel to the SiC – Si interface read 1.62 Å in SiC, corresponding to {2 2 0} planes of 3C-SiC, and 1.92 Å in Si, corresponding to {2 2 0} planes of Si. Consequently, the SiC layer is formed by cubic silicon carbide (3C-SiC), known for its zinc-blende crystal lattice, and reveals the same symmetry of the atomic structure as the Si substrate, possessing smaller interatomic distances.

The SIMS and TEM results are consistent with those of the preceding sections. The persistence of a substrate Si signal in XPS at 240 min (Section 3.1), the retained surface roughness (Section 3.2), and the broad SiC/Si transition observed here collectively support the formation of a continuous but laterally non-uniform SiC layer bounded by a rough interface. Such a sustained SiC formation after initial surface coverage requires a transport of Si from the substrate to the reaction front, as first established for this reaction system by Mogab and Leamy [25], and subsequently confirmed through studies of void and micropipe formation at the SiC/Si interface [7,26,27].

## 4. Conclusion

This study demonstrates the direct, time-resolved formation of SiC on Si(111) by $C_2H_4$ exposure at 800 °C under UHV, from the earliest nucleation stage to an extended layer, using XPS, correlative SEM/AFM, site-specific AES, and SIMS combined with STEM/EELS. XPS shows a systematic chemical conversion with exposure time: the Si–C component of the Si 2p spectrum rises from a small fraction at short exposure to a plateau of 80–81% at the longest exposure times, while a residual elemental Si signal of about 19–20% persists even at these longest exposures rather than disappearing entirely. This persistence of unconverted Si, rather than its complete elimination, indicates that the reaction saturates and becomes self-limiting rather than proceeding to a fully closed, homogeneous layer. Correlative SEM and AFM show that this chemical evolution is accompanied by a parallel morphological one: discrete islands nucleate at short exposure times and progressively coalesce into a more continuous surface as exposure continues, consistent with an island-mediated growth mode that gradually approaches full coverage, however accompanied with the appearance of larger voids. Site-specific AES on the earliest-stage sample directly ties this morphology to the underlying chemistry. The C/Si ratio at an island was consistently 2.0–2.3 times higher than at the surrounding substrate, both before and after two $Ar^+$ sputtering cycles. This shows carbon at the islands in a strongly bound configuration, providing direct, spatially resolved evidence that these islands are sites of early-stage carbide nucleation. SIMS depth profiling and STEM cross-sectional imaging indicates a broad, non-abrupt SiC/Si transition. The STEM image of a homogeneous part selected in a SiC island – grain shows a crystalline epitaxial layer having sharp interfaces both with the Mo covering layer and Si substrate. The thickness of the layer was 9 nm which corresponds to the width of the maximum $C^+$ SIMS signal. The STEM-EELS measurements have confirmed the Si-C nature of the layer and determined interplanar spacings corresponding to the cubic 3C-SiC polytype.

Together, the techniques converge on a single, internally consistent picture: SiC nucleates as discrete, carbon-enriched islands that grow and coalesce with continued $C_2H_4$ exposure, forming a layer gradually approaching a full coverage frustrated by the appearance of larger voids; the resulting SiC/Si interface is generally rough and gradually varying, atomically abrupt only in some parts of islands - grains. The knowledge acquired during this study will be utilized in an application of SiC as the buffer layer for a direct growth of $Mo_2C$ MXenes.

## Acknowledgement

We acknowledge the support by the Czech Science Foundation (grant No. 23-07617S), OP JAK (project No. CZ.02.01.01/00/22_008/0004594 TERAFIT), and CzechNanoLab Research Infrastructure (ID 90251), funded by MEYS CR, is gratefully acknowledged for the financial support of the measurements. The authors acknowledge Dr V. Danchuk for the nano-SAM measurements.